\documentclass[%
showpacs,
amsmath,amssymb,
aps,
]{revtex4-1}

\usepackage[dvipdfmx]{graphicx,color}
\usepackage{bm}
\usepackage{setspace}

\begin{document}

\preprint{}

\title{A voter model on networks  and 
  multivariate beta distribution}

\author{Shintaro Mori}
\email{shintaro.mori@hirosaki-u.ac.jp}
\affiliation{
Department of Mathematics and Physics,
Faculty of Science and Technology, 
Hirosaki University, \\ 
Bunkyo-cho 3, Hirosaki, Aomori 036-8561, Japan
}

\author{Masato Hisakado}
\email{hisakadom@yahoo.co.jp}
\affiliation{
  Nomura Holdings Inc., \\ 
  Otemachi 2-2-2, Chiyoda-ku, Tokyo 100-8130, Japan
}%

\author{Kazuaki Nakayama}
\email{nakayama@math.shinshu-u.ac.jp}
\affiliation{
Department of Mathematical Sciences,
Faculty of Science, Shinshu University, \\
Asahi 3-1-1, Matsumoto, Nagano 390-8621, Japan
}%

\date{\today}

\begin{abstract}
  In elections, the vote shares or turnout rates
  show
  a strong spatial correlation.
  The logarithmic decay with distance suggests that a
  2D noisy diffusive equation describes the system.
  Based on the study of U.S. presidential elections data,
  it was determined that the fluctuations of
  vote shares also exhibit a strong and long-range
  spatial correlation. Previously, it was considered difficult to
  induce strong and long-range spatial correlation of the
  vote shares
  without breaking
  the empirically observed narrow distribution. 
  We demonstrate that 
  a voter model on networks shows such a behavior.  
  In the model, there are many voters in a node who are affected by
  the agents in the node and by the 
  agents in the linked nodes.
  A multivariate Wright-Fisher diffusion equation for the joint
  probability density of the vote shares is derived.
  The stationary distribution 
  is a multivariate generalization of the beta
  distribution. In addition, we also estimate the equilibrium values and
  the covariance matrix of the vote shares and obtain a correspondence
  with a multivariate normal distribution.
  This approach largely simplifies the calibration of
  the parameters in the modeling of elections. 
\end{abstract}

\maketitle


\section{\label{sec:intro}Introduction}
Social physics has become an active
research field\cite{Galam:2008,Castellano:2009,Ormerod:2012,Pentland:2014}
and many studies have been devoted to the understanding of
social phenomena and interacting human behaviors
\cite{Kirman:1993,Lux:1995,Cont:2000,Salganik:2006,Watts:2007,Conradt:2009,Rendell:2010,Bentley:2011,Bond:2012,Mori:2010,Mori:2012,Mori:2016,Nakayama:2017,Hisakado:2018}.
Opinion dynamics is a central research theme,
and empirical studies based on election data have been extensively pursued 
\cite{Araujo:2010,Borghesi:2010,Borghesi:2012,Gracia:2014}.
In these investigations, the correlation between the voters' decisions was evaluated
by studying the dependence of 
the variance of the turnout rate on the
number of voters $N$\cite{Borghesi:2010,Borghesi:2012}. 
If the voters' decisions are independent, the variance of the turnout rate
should be proportional to $N^{-1}$.
An empirical study of French election data showed that
the voters' decisions were proportional to the power of $N^{-3/4}$.
In addition, it was determined that 
the spatial correlation of the turnout rate in each election
exhibited a logarithmic decay with distance that
suggested a description based on a 2D noisy diffusion equation.

A threshold model was introduced for the binary decision of an individual
with intension field\cite{Borghesi:2010}. If the intension of an
individual exceeds a certain
threshold, the decision is one. When it is below the threshold, the decision
is 0. The intension field was decomposed into the sum of a noise which
is an instantaneous contribution, a space
dependent "cultural" field and the influence of the
decision of other individuals.
Here, "cultural field" encodes all the local, stable features that
influence the final decision.
Without the noise and the cultural field, the model simplifies to the Random
Field Ising Model\cite{Galam:1982,Galam:1991,Galam:1997}.
It was concluded that the long-range spatial correlations cannot be due to
the influence of the decision of others, because the interaction cannot
induce the empirically observed
unimodal and narrow distribution of turnout rates.
The long-range spatial correlation was thus attributed
to that
of the "cultural field".
As a phenomenological model of the cultural field, a 2D noisy diffusion
equation was proposed.

The voter model and its noisy extension 
have been studied extensively in opinion
dynamics\cite{Liggett:2005,Mobilia:2003,Suchecki:2005,
  Mobilia:2007,Sood:2005,Sood:2008,
  Castellano:2009,Carro:2016}.
In particular, the validity of the voter model as a model for elections
was tested in the U.S. presidential election\cite{Gracia:2014}.
In this model, agents move between their living places and their workplaces. 
In both places, their decisions are affected by other voters.
The model is called the social influence recurrent mobility (SIRM) model.
Based on the diffusion approximation of the model,
a noisy diffusion equation was derived. By balancing the strength of
the noise with the voter model’s
consensus mechanism or the force of conformity, it was concluded that
the SIRM model can reproduce the statistical features of the vote-share
in presidential elections, i.e. the stationarity of the variance of
vote-share distributions and the long-range spatial correlation
that decays logarithmically with distance.
However, the model has a drawback in that under certain circumstances,
the noise might break the
range of vote shares.  This was addressed by
introducing  the 
beta distributed noise\cite{Michaud:2018}.
Furthermore, 
a generalization to the case of more than two political parties was also
proposed in the same framework.

In this report, we study the
correlation of the
fluctuations of vote shares using theoretical and empirical methods.
Based on U.S. presidential election data, we show that the correlation
of the fluctuation of the vote shares between the nearest
neighbor counties exceed 80\% and it is much higher than that
of  the temporal averages of the vote shares. 
 Furthermore, as with the  latter ones ,
the fluctuation also shows long-range spatial correlation.
 In the threshold model without the influence of the decisions of others,
the fluctuations of the vote shares are independent of each other
even if the cultural field shows a strong spatial correlation.
The correlation of the cultural field only affects the correlation of
 the temporal averages of the vote shares. 
The threshold model with the social influence term  
is inappropriate for inducing such a strong correlation of the fluctuations
because it contradicts the empirical results.
Therefore, an alternate model that can incorporate a strong correlation
without losing the unimodality of the vote share distribution
should be introduced.
According to the results of the SIRM model, a voter model should be a
good candidate.
We show that the vote shares of a voter model on networks
obeys a multi-variate beta distribution which 
can incorporate strong correlation without losing the
unimodality of the distribution of the vote shares.
Furthermore, the distribution is similar to the multivariate normal
distribution and the calibration of the model parameters is easy. 

The paper is organized into multiple sections.
In Sec.~\ref{sec:data}, the U.S. presidential election data is studied
and the vote shares are decomposed into the equilibrium values and
the fluctuations around them.
The  cultural field  
are encoded in the former and both
exhibit strong and long-ranged spatial correlation.
It is shown that the vote shares approximately obey a multivariate normal
distribution.
A voter model on networks is introduced 
in Sec.~\ref{sec:model}. The multivariate
Wright-Fisher diffusion equation is then derived for the joint probability
density function (pdf) of the vote shares. The stationary distribution is a
multivariate beta distribution. We approximate the distribution
using a multivariate normal distribution and estimate the covariance matrix
of the vote shares.  Sec.~\ref{sec:simulation} is devoted to
the numerical analysis and verification of the theoretical results.
Sec.~\ref{sec:Con} includes the conclusions and
discussions of future problems.

\section{\label{sec:data} Empirical study}

U.S. presidential election data from 1980 to 2016 were studied.
A total of ten elections occurred during this interval and they are labeled as
$t=1,2\cdots,T=10$ where $t=1$ corresponds to the election in 1980.
The data of 3105 counties was studied
and label as $i=1,2,\cdots,I$.
The data consist of the number of votes $N(i,t)$
and the number of votes for the democratic party $n(i,t)$
in county $i$ and election $t$.
The total number of votes cast in election $t$ is calculated
as $N_{T}(t)=\sum_{i}N(i,t)$.
The votes that were not cast for
either the Democratic party or the Republican party were excluded, and
the votes for the latter party is given by $N(i,t)-n(i,t)$.
We denote the vote share for the democratic party as
$v(i,t)\equiv n(i,t)/N(i,t)$.

\begin{figure}[htbp]
\begin{center}
\begin{tabular}{cc}
\includegraphics[width=8cm]{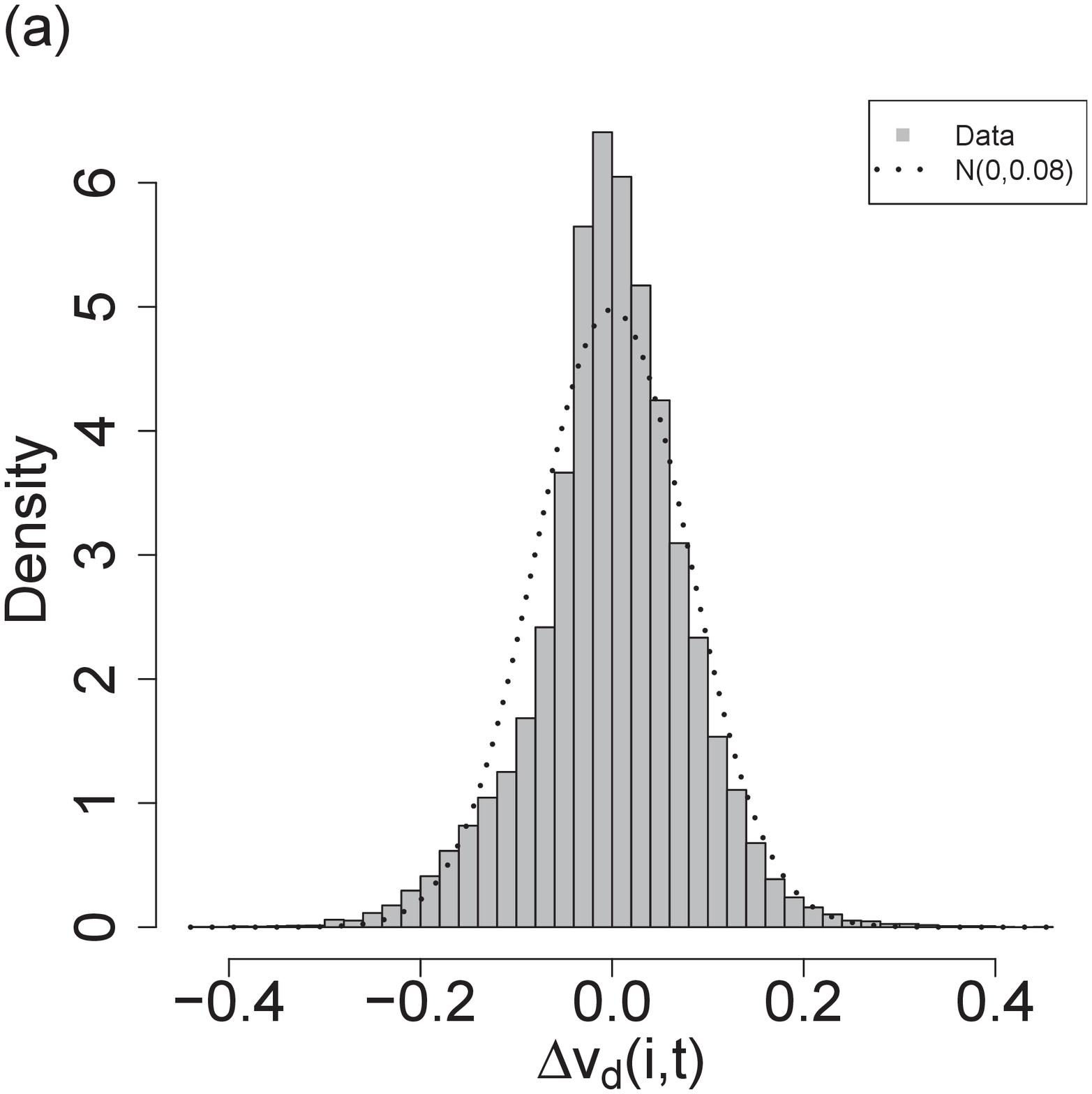} &  
\includegraphics[width=8cm]{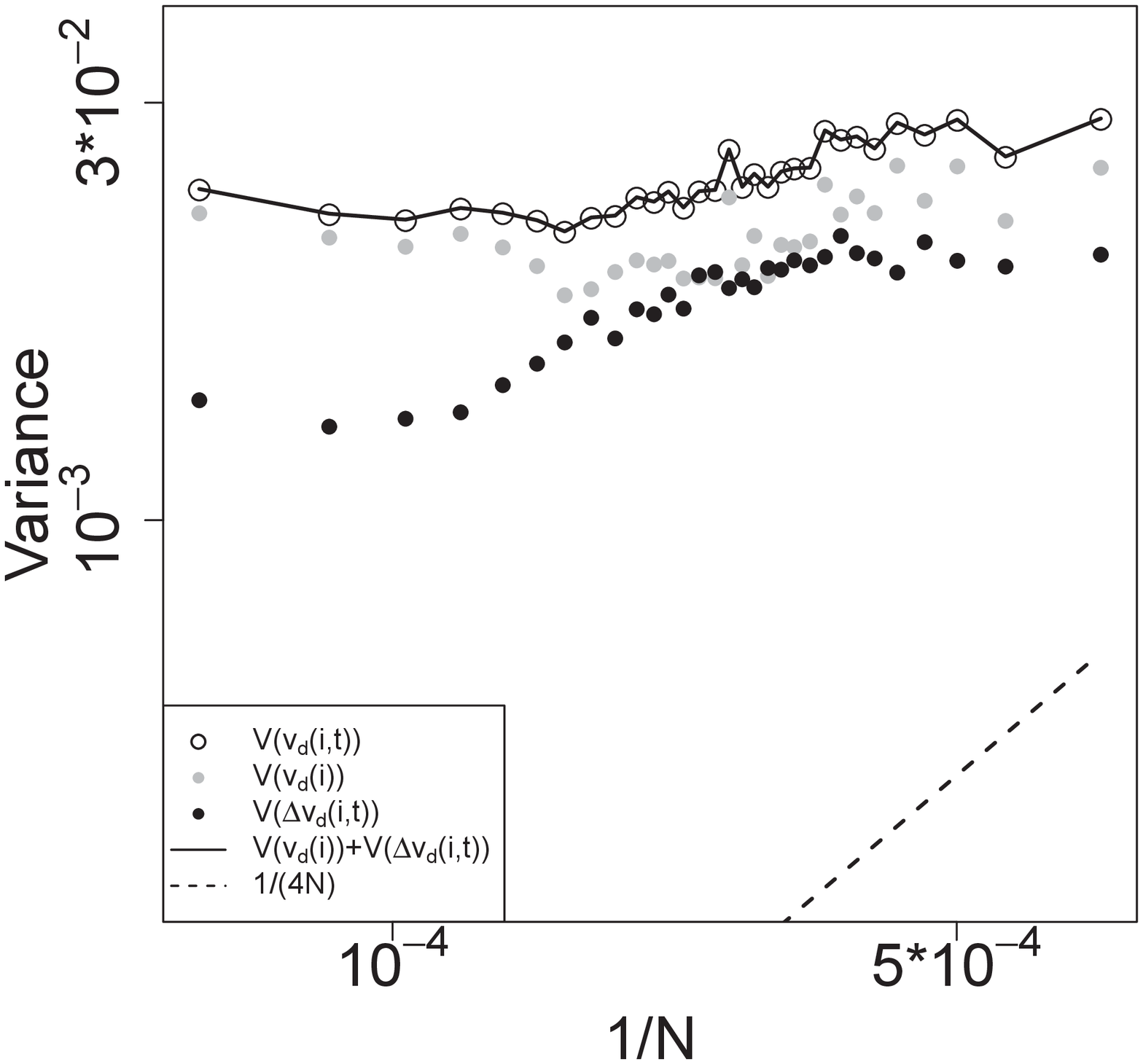}  
\end{tabular}
\end{center}
\caption{\label{fig:Vvd}
  (a) Plot of the distribution of $\Delta v_{d}(i,t)=v_{d}(i,t)-v_{d}(i)$.
  The dotted line shows the normal distribution with the same mean and
  the variance.
  (b)
  Plot of the variance of $v_{d}(i,t)$ and its decomposition to
  the variance of $v_{d}(i)$ and $\Delta v_{d}(i,t)$ vs. the
  inverse of the average
  value of $N(i)$. The solid line shows the sum of the two variances.
  The broken line shows $1/(4N)$.
}
\end{figure}

Initially, we detrend the vote share data.
The weighted spatial average of $v(i,t)$ is estimated as
\[
v(t)\equiv \sum_{i}n(i,t)/N_{T}(t).
\]
We estimate the temporal average of $v(t)$ as $v_{avg}\equiv
\sum_{t}v(t)/T$ and
obtain the detrended vote share as
\[
v_{d}(i,t)\equiv v(i,t)-(v(t)-v_{avg}).
\]
Based on this process, the weighted spatial average of $v_{d}(i,t)$
becomes $v_{avg}$ and it does not depend on $t$.
The temporal average of $v_{d}(i,t)$ is defined
as $v_{d}(i)=\sum_{t}v_{d}(i,t)/T$, which is an
estimate of the equilibrium values of $v_{d}(i,t)$
in county $i$. 
We interpret $v_{d}(i)$ as the "cultural field" because it reflects
the local and stable features. 
We denote the fluctuation (deviation)
of $v_{d}(i,t)$ around $v_{d}(i)$ as
$\Delta v_{d}(i,t)\equiv v_{d}(i,t)-v_{d}(i)$.
Figure \ref{fig:Vvd}(a) shows the distribution
of $\Delta v_{d}(i,t)$. The standard deviation(SD)
is approximately 8\% and slightly left-skewed.

We study the $N$ dependence of the variance of $v_{d}(i,t)$.
The spatial average of $v_{d}(i)$ is denoted as
$v_{d,avg}=\sum_{i}v_{d}(i)/I$. 
The fluctuation of $v_{d}(i,t)$ around
$v_{d,avg}$ is decomposed as the sum of the fluctuation of
$\Delta v_{d}(i,t)$ and that of $v_{d}(i)$.
\begin{eqnarray}
\mbox{V}(v_{d}(i,t))&=&\frac{1}{IT}\sum_{i,t}(v_{d}(i,t)-v_{d,avg})^2=
  \frac{1}{IT}\sum_{i,t}(v_{d}(i,t)-v_{d}(i)+v_{d}(i)-v_{d,avg})^2  \nonumber \\
  &= &\frac{1}{IT}
  \sum_{i,t}\{\Delta v_{d}(i,t)^2+(v_{d}(i)-v_{d,avg})^2\}  \nonumber \\
  &=&\mbox{V}(\Delta v_{d}(i,t))+\mbox{V}(v_{d}(i)) \nonumber
\end{eqnarray}

As $\sum_{t}\Delta v_{d}(i,t)=0$, the cross term vanishes and 
the third equality holds. 
The $N$ dependence of the
fluctuation of $\Delta v_{d}(i,t)$ is then investigated.
We bin $v_{d}(i,t)$ according to $N(i)\equiv \sum_t N(i,t)/T$
into 31 classes and each class contains 100
counties, with almost the same number of average votes $N(i)$.
As previously discussed \cite{Borghesi:2010}, if the voters choose
independently, the variance of $\Delta v_{d}(i,t)$ is proportional
to the inverse of $N(i)$ as $v_{d}(i)(1-v_{d}(i))/N(i)$.
Figure \ref{fig:Vvd}(b) plots 
V$(v_{d}(i,t))$, V$(v_{d}(i))$ and
V$(\Delta v_{d}(i,t))$ vs. $1/N$.
It is evident that, V$(\Delta v_{d}(i,t))$ is much
larger than $1/4N$ in all the bins.
One also observes that the decomposition of the variance holds.

Next, we study the spatial correlation of $v_{d}(i,t)$.
There are $I(I-1)/2\simeq 4.81\times 10^6$ pairs of counties
$(i,j),i,j\in \{1,2,\cdots,I\}$.
They are sorted according to the distance $r(i,j)$ between county $i$ and
county $j$.
The database of the national bureau of economic research
is utilized to obtain information on the inter-county distance\cite{NBER}.
The distance is taken as the separation of the centroids. 
The sorted pairs $(i,j)$ are binned into 481 classes and each class
contains $10^4$ pairs. 
We label the bin of the county pairs $(i,j)$ separated by
their average distance $r$ as $R(r)$ and $|R(r)|=10^4$ 
represents the number of pairs in the bin.

\begin{figure}[htbp]
\begin{center}
\begin{tabular}{cc}
\includegraphics[width=8cm]{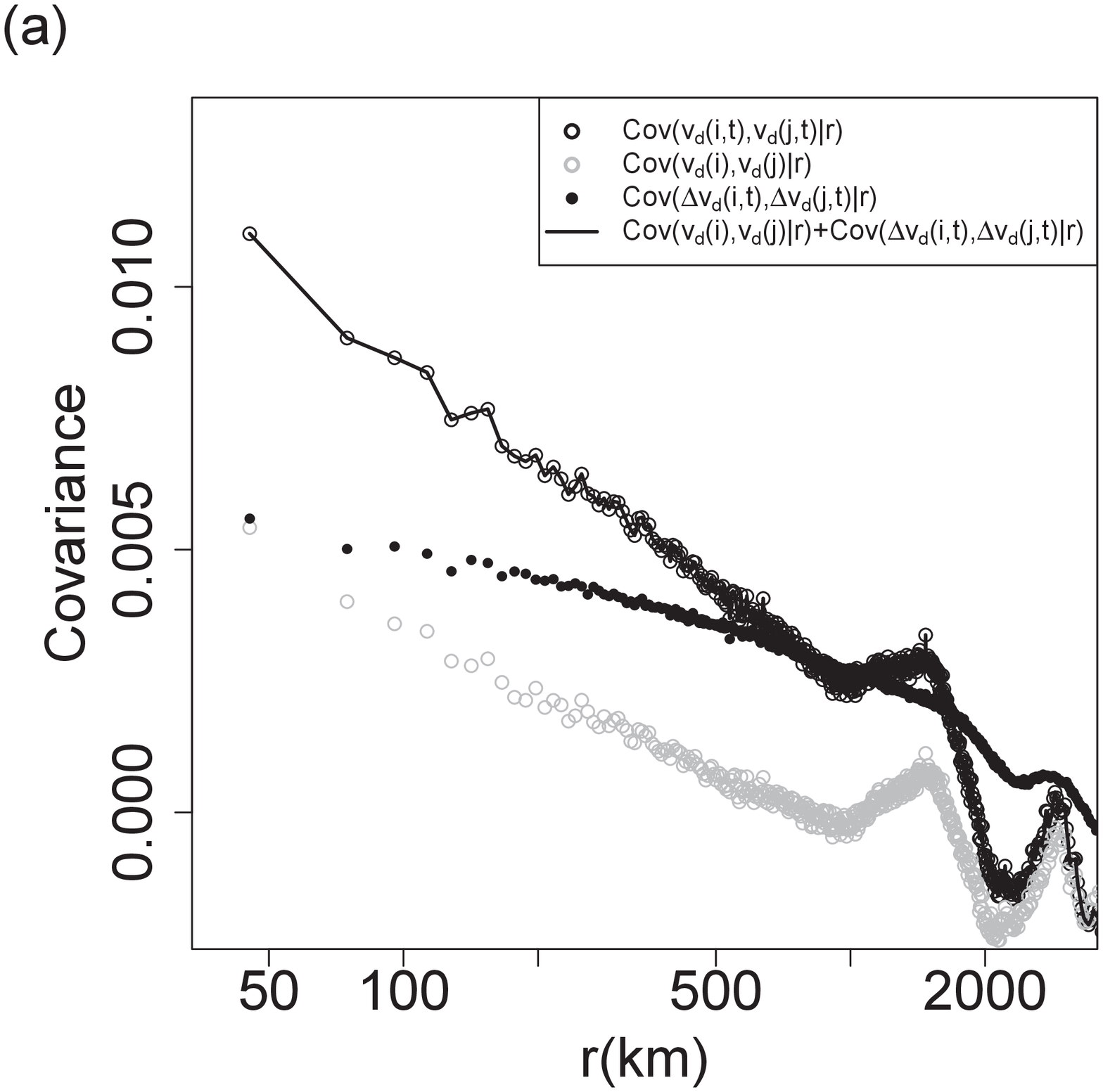} &  
\includegraphics[width=8cm]{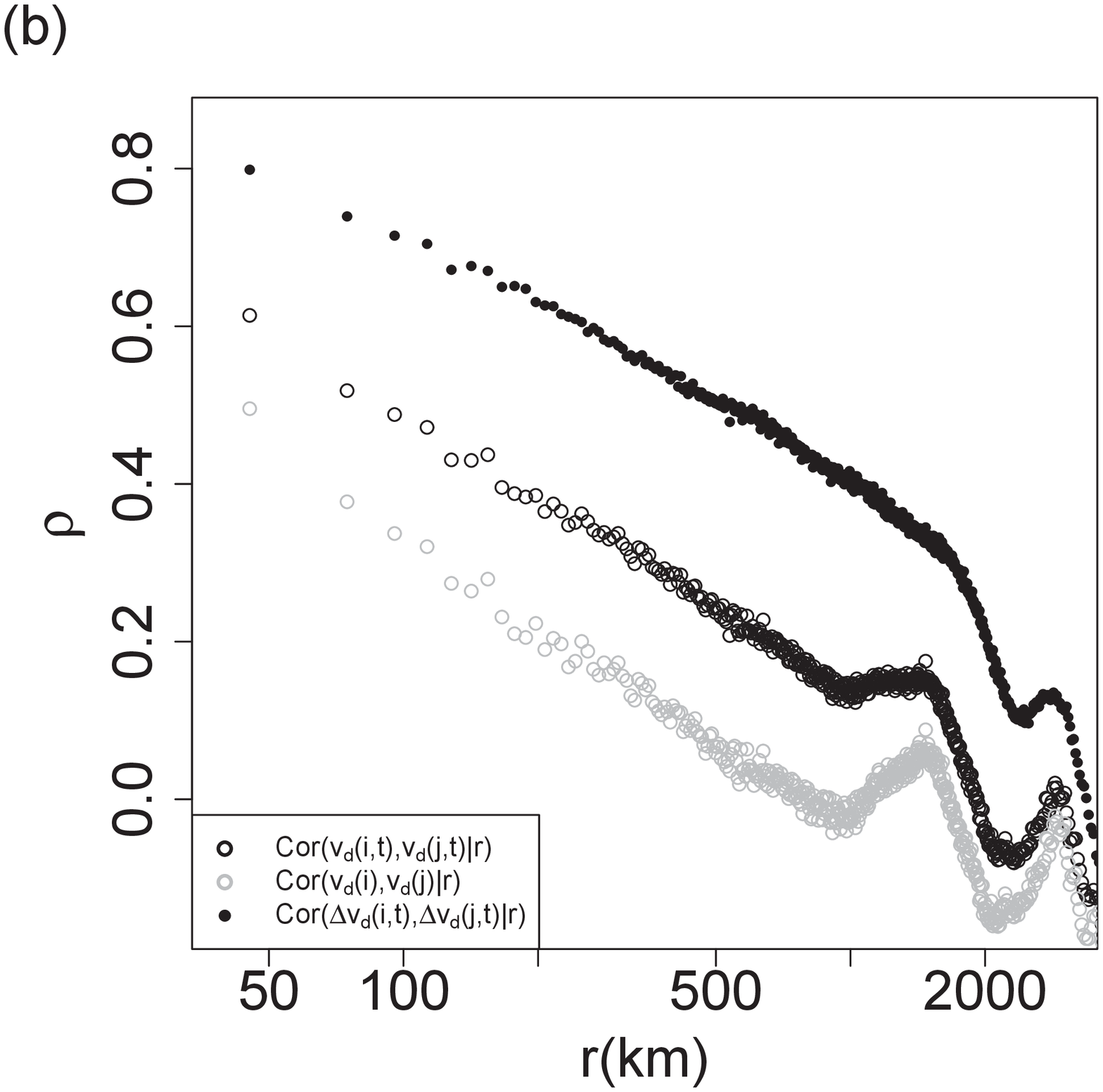}  
\end{tabular}
\end{center}
\caption{\label{fig:r_vs_Cor}
Semi-logarithmic plots of the (a) covariance and (b) correlation
of $v_{d}(i,t),v_{d}(i)$ and $\Delta v_{d}(i,t)$ vs. $r$.
The solid line in the left figure shows the sum of the covariance
of $v_{d}(i)$ and $\Delta v_{d}(i,t)$. According to the decomposition of the
covariance, the contribution of 
$\Delta v_{d}(i,t)$ is larger than that of $v_{d}(i)$.
The correlation coefficient of $\Delta v_{d}(i,t)$ is also
larger than that of $v_{d}(i)$.
}
\end{figure}

The covariance of $v_{d}(i,t)$ and
$v_{d}(j,t)$ of the pairs in  $R(r)$
are defined as:
\[
\mbox{Cov}(v_{d}(i,t),v_{d}(j,t)|r)=\frac{1}{T}\sum_{t}
\frac{1}{|R(r)|}\sum_{(i,j)\in R(r)}
(v_{d}(i,t)-v_{d}(R(r)))(v_{d}(j,t)-v_{d}(R(r))).
\]
Here, 
$v_{d}(R(r))$ is the average value of $v_{d}(i,t),v_{d}(j,t)$ in $R(r)$.
\[
v_{d}(R(r))=\sum_{t}\sum_{(i,j) \in R(r)}v_{d}(i,t)/|R(r)|T
=\sum_{t}\sum_{(i,j) \in R(r)}v_{d}(j,t)/|R(r)|T.
\]
The covariance is then decomposed using the following identity:
\begin{eqnarray}
&&(v_{d}(i,t)-v_{d}(R(r)))(v_{d}(j,t)-v_{d}(R(r)))  \nonumber \\
&=&(v_{d}(i,t)-v_{d}(i)+v_{d}(i)-v_{d}(R(r)))(v_{d}(j,t)-v_{d}(j)+v_{d}(j)-v_{d}(R(r))) \nonumber
\end{eqnarray}
We then obtain the next decomposition of the covariance as the cross term
vanishes by the equality
$\sum_t(v_{d}(i,t)-v_{d}(i))=0$. 
\[
\mbox{Cov}(v_{d}(i,t),v_{d}(j,t)|r)= 
\mbox{Cov}(\Delta v_{d}(i,t),\Delta v_{d}(j,t)|r)+\mbox{Cov}(v_{d}(i),v_{d}(j)|r).
\]
 By normalizing the covariances with the variances,
  we estimate the correlation coefficients $\rho$ for $v_{d}(i,t),v_{d}(i)$
  and $\Delta v_{d}(i,t)$. Figure \ref{fig:r_vs_Cor}
  represents the semi-logarithmic plot of the covariance and 
  the correlation vs. $r$.

The left figure plots the covariance vs. $r$. The solid line shows
the sum of the covariance of $v_{d}(i)$ and $\Delta v_{d}(i,t)$.
It is evident that the sum lies on the plot of the covariance of $v_{d}(i,t)$.
The covariance of $\Delta v_{d}(i,t)$ is larger than that of $v_{d}(i)$.
The right figure plots the correlation coefficient $\rho$ vs. $r$.
In all the three cases, the correlation exhibits a logarithmic decay with $r$. 
The interesting point is that the correlation of $\Delta v_{d}(i,t)$
is the largest and it is over 80\% for the bin of the nearest
neighbor county pairs.
The correlation of $\Delta v_{d}(i,t)$ for nearest neighbor pairs
is estimated to be $83.4\%$. The spatial correlation of $v_{d}(i)$ implies
that the cultural field of two counties are similar when these counties are
near each other. The spatial correlation of $\Delta v_{d}(i,t)$ represents the
co-movement of the voters' decisions in the two counties.
These two correlations have completely different physical origins. 
For $r\ge 10^3[km]$, an unusual behavior is observed.
The correlation decays with $r$ and
local maxima appear in the correlation of $v_{d}(i)$ and $v_{d}(i,t)$.
The correlation of $\Delta v_{d}(k,t)$ shows a monotonically decreasing
behavior up to 2000[km].

The results will now be summarized.
We decompose $I$ variables 
$\vec{v}_{d}(t)=(v_{d}(1,t),\cdots,v_{d}(I,t))$
as the sum of $\vec{v}_{d}=(v_{d}(1),\cdots,v_{d}(I))$ and
$\Delta \vec{v}_{d}(t)=(\Delta v_{d}(1,t),\cdots,\Delta v_{d}(I,t))$.
$\vec{v}_{d}(t)$ fluctuates around $\vec{v}_{d}$.
$\vec{v}_{d}$ is a proxy for the cultural field   
and  $\Delta \vec{v}_{d}(t)$ shows a stronger spatial correlation
than $\vec{v}_{d}$.
$\vec{v}_{d}(t)$
approximately obeys a
multivariate normal distribution with a mean of
$\vec{v}_{d}$
and the variance-covariance matrix
V$(\vec{v}_{d})$+V$(\Delta \vec{v}_{d}(t))$.
\[
\vec{v}_{d}(t)=\vec{v}_{d}+\Delta \vec{v}_{d}(t)
\sim \mbox{N}_{I}(\vec{v}_{d},
\mbox{V}(\vec{v}_{d})+\mbox{V}(
\Delta \vec{v}_{d}(t)))
\]
It should be noted that the time scale of $\Delta \vec{v}_{d}(t)$
and that of $\vec{v}_{d}$ are quite different. 
The voting habits of the different regions
is extremely persistent
and the time scale of $\vec{v}_{d}$ is a century or more\cite{Borghesi:2010}.
However, $\Delta \vec{v}_{d}(t)$ fluctuates rapidly 
and the time scale is short.

\section{\label{sec:model} Social influence model on networks}

\begin{figure}[htbp]
\begin{center}
\includegraphics[width=12cm]{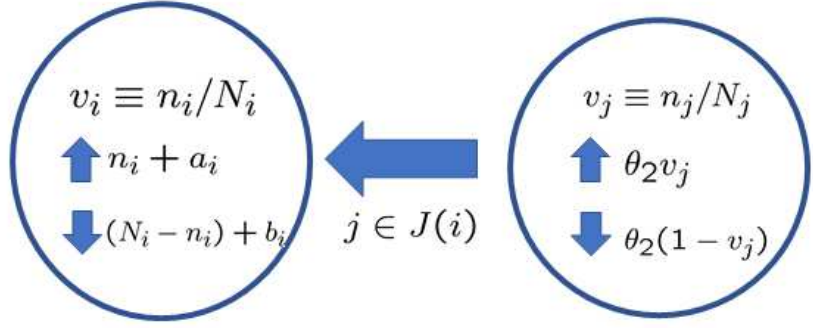}
\end{center}
\caption{\label{fig:model}
Illustration of the voter model.
The big circles represent node $i$ and
node $j\in J(i)$ that is linked with node $i$.
There are $N_i$ and $N_j$ voters in node $i$ and node $j$,
respectively. Among them $n_i(n_j)$ voters
(UP arrow) and $N_i-n_i(N_j-n_j)$ voters (Down arrow)
are for and against an option in node $i(j)$.
Initially, a susceptible voter is chosen at random from 
node $i$.
Next, an infectious voter is chosen from node $i$ or node $j\in J(i)$.
Node $i$ has $N_i-1$ infectious voters after a voter is chosen from the node.
In addition, there are $\theta_1=a_i+b_i$ voters who do not change their
decisions. Among them,  $a_i$ and $b_i$ voters are for and
against the option, respectively, $\mu_i=a_i/\theta_1$ is the intrinsic tendency
of node $i$ to vote for the option and  $\theta_1$ controls
the strength of the tendency. 
Node $j$ has $\theta_2$ infectious voters.}
\end{figure}

We now introduce a voter model on networks.
There are $I$ nodes and they are labeled as $i=1,2,\cdots,I$.
The link set $E=\{(i,j)\}$ consists of links that connect node $i$ and $j$.
$J(i)\equiv \{j|(i,j)\in E\}$ denotes the set of nodes that are linked with
node $i$ and 
$|J(i)|$ is the number of nodes linked with node $i$. 
In each node, there are $N_i$ agents 
whose decisions obey the dynamics
of the voter model\cite{Liggett:2005,Gracia:2014}.
One agent is chosen at
random from $N_{T}=\sum_i N_i$ agents.
If the agent is from node $i$, another agent is chosen from node $i$
or from node $j\in J(i)$, which is connected to node $i$.
$n_i$ and $v_i\equiv n_i/N_i$ denote the number of votes and the vote share
of an option. We assume that the intrinsic tendency of the voters
in node $i$ to vote for
an option is determined by the parameters $\mu_i$ and $\theta_1$.
$\mu_i$ is the probability that a voter votes for an option and
$\theta_1$ is a parameter that controls
the variance of the vote share.  Intuitively, $\theta_1$
represents the number of voters who are not influenced by other
voters.
$a_i\equiv \mu_i\theta_1$ and $b_i\equiv (1-\mu_i)\theta_1$
corresponds to the number of such voters who choose and do
not choose the option, respectively.
The strength of the influence of the voters in the linked node
$j\in J(i)$ is denoted as $\theta_2$. 
Figure \ref{fig:model} illustrates the model.

The probability that the number of voters for an option $n_i$
increases by 1
is written as the product of the probabilities 
of the next two processes. Initially, a voter of node $i$ 
who does not choose the option
is selected. The probability is $(N_i-n_i)/N_{T}$.
Secondly, an infectious voter who can affect the voter and
choose the option is selected. The infectious voters
should live in node $i$ or in the linked node
$j,j\in J(i)$. $\theta_1$ voters in node
$i$ also can affect the voter. The number of infectious voters in
the linked node $j$ is set to be $\theta_2$.
This is the simplification of the infectious process.
The total number of the infectious voter in node $i$
is $N_i-1+\theta_1+\theta_2\cdot |J(i)|$.
Here $-1$ of  $N_i-1$ indicates that we omit the voter who
is chosen in the first process.
Among them $n_i+a_i+\theta_2\cdot \sum_{j\in J(i)} v_j$ choose the option.
The probability for the second process is then:
\[
\frac{n_i+a_i+\theta_2\cdot \sum_{j\in J(i)}v_j}{N_{i}-1+\theta_1+\theta_2\cdot |J(i)|}.
\]
The probability for $n_i\to n_i+1$ is then written as:
\[
P(n_{i}\to n_{i}+1)=\frac{N_i-n_i}{N_{T}}\cdot
\frac{n_i+a_i+\theta_2\cdot \sum_{j\in J(i)}v_{j}}{N_{i}-1+\theta_1+\theta_2\cdot |J(i)|}.
\]

Likewise, the probability that $n_{i}$ decreases by 1 is written as:
\[
P(n_{i}\to n_{i}-1)=\frac{n_i}{N_{T}}\cdot
\frac{N_i-n_i+b_i+\theta_2\cdot \sum_{j\in J(i)}(1-v_{j})}
{N_{i}-1+\theta_1+\theta_2 \cdot |J(i)|}.
\]
There are two main differences in this model
compared to the SIRM model\cite{Gracia:2014,Michaud:2018}.
One difference is the node dependent terms $\mu_i$  
and $\theta_1$. If the influence from other
voters in the linked nodes $j\in J(i)$ is turned off
by setting $\theta_2=0$, the model reduces to Kirman's ant
colony model\cite{Kirman:1993,Hisakado:2015}.
 The stationary probability distribution of $n_{i}$ is the beta-binomial
 distribution.
 The vote share $v_i$ obeys a
 beta distribution with shape parameters $(a_i,b_i)$
 in the limit $N_i\to \infty$.
  The derivation of the beta distribution is
 given in Appendix B. .
 \[
v_i \sim \mbox{Beta}(a_i,b_i)
\]
 The expectation value of $v_i$ is
 $\mu_i$ and the variance of $v_i$ is $\mu_i(1-\mu_i)/(\theta_1+1)$.
 This variance originates from the interaction between the voters in node $i$.
 The correlation of the voters’ binary choices is
 $1/(\theta_1+1)$\cite{Hisakado:2006}.
 The second change is the normalization of $n_j$ by $N_j$  and
 we use $v_j=n_j/N_j$.
 The mathematical reason for the modification is to avoid the ill-posedness
 in the original SIRM model\cite{Michaud:2018}. As we shall show shortly,
 if we normalize as indicated, the noise term becomes proportional
 to $\sqrt{v_i(1-v_i)}$ as in the Wright-Fisher diffusion equation 
 and it does not break the condition
 $v_i \in (0,1)$ even when $v_i$ approaches 0 or 1.
 
 The raising operator $R_{i}\equiv P(n_i\to n_i+1)$ is written
 using vote shares $\vec{v}=(v_1,v_2,\cdots,v_I)$ as:
 \[
 R_{i}(\vec{v})=
 \frac{N_{i}}{N_{T}}\cdot (1-v_i)\cdot \frac{v_i+a_i/N_i
   +\theta_2\cdot \sum_{j\in J(i)}v_j/N_i}{1-1/N_i+\theta_{1}/N_i+\theta_2\sum_{j\in J(i)}/N_i}.
 \]
 The lowering operator $L_{i}\equiv P(n_{i}\to n_{i}-1)$ is also written as:
 \[
 L_{i}(\vec{v})=
 \frac{N_{i}}{N_{T}}\cdot v_i\cdot \frac{(1-v_i)+b_i/N_i
   +\theta_2\cdot \sum_{j\in J(i)}(1-v_j)/N_i}{1-1/N_i+\theta_{1}/N_i+\theta_2\sum_{j\in J(i)}/N_i}.
 \]
 We write $v_{i,c}$ for the average value of the vote shares $\{v_{j}\}$ of
 the linked nodes $j\in J(i)$. 
 \[
v_{i,c}=\sum_{j\in J(i)}v_{j}/|J(i)|.
\]
 $R_{i},L_{i}$ are then rewritten as
 \begin{eqnarray}
 R_{i}(\vec{v})&=&
 \frac{N_{i}}{N_{T}}\cdot (1-v_i)\cdot \frac{v_i+a_i/N_i
   +\theta_2\cdot |J(i)| \cdot v_{i,c}/N_i}
      {1-1/N_i+\theta_{1}/N_i+\theta_{2}\cdot |J(i)|/N_i}, \nonumber \\
 L_{i}(\vec{v})&=&
 \frac{N_{i}}{N_{T}}\cdot v_i\cdot \frac{(1-v_i)+b_i/N_i
   +\theta_{2}\cdot |J(i)|\cdot (1-v_{i,c})/N_i}
      {1-1/N_i+\theta_{1}/N_i+\theta_{2}\cdot |J(i)|/N_i}. \nonumber       
 \end{eqnarray}

 The stochastic differential equation \cite{Gardiner:2009} for $v_i$ is
 written with drift $d_i$ and
 diffusion $D_{i}$ as:
 \[
d v_{i}=d_{i}dt+\sqrt{D_{i}}dW_{i}(t).
\]
Here, $dW_i(t)$ is iid white noise, or Brownian motion.
The drift term $d_i$ is estimated as:
\[
d_i=\frac{\delta v_i}{\delta t}(R_i-L_i)
=\frac{\delta v_i}{\delta t}\frac{1}{N_T}
(a_i-\theta_{1}v_i+\theta_{2}\cdot |J(i)|\cdot (v_{i,c}-v_{i})).
\]
Here, we take the limit $N_i\to \infty$ in the second equality.
The diffusion term $D_i$ is estimated as:
\[
D_{i}=\frac{(\delta v_i)^2}{\delta t}(R_i+L_i)
=\frac{(\delta v_i)^2}{\delta t}\frac{N_i}{N_T}2v_i(1-v_i).
\]
If we set $N_i=N$ and $N_{T}=IN$, we have $\delta v_i=1/N$ and $\delta t=1/IN^2$.
$d_i$ and $D_i$ are written as:
\begin{eqnarray}
d_i&=&(a_i-\theta_{1}v_i+\theta_{2}\cdot |J(i)|\cdot (v_{i,c}-v_{i})) \nonumber \\
D_i&=&2v_i(1-v_i). \nonumber
\end{eqnarray}
The Fokker-Plank equation for the time evolution of the
joint probability density function $f(\vec{v},t)$ is give as:
\begin{equation}
\partial_{t}f(\vec{v},t)=-\sum_{i}
\left\{\partial_{i}d_i-\frac{1}{2}\partial_{i}^2D_i   \right\}f . \label{eq:FP}
\end{equation}
Here, we write the derivative by $v_i$ as $\partial_i$.
This is a multi-variate Wright-Fisher diffusion
process \cite{Ethier:1986}.

Given that the drift and diffusion terms
do not explicitly depend on $t$, the stochastic
system is a statistically stationary process and
the solution of the Fokker-Planck equation converges to a
stationary distribution\cite{Gardiner:2009}.
\[
f_{st}(\vec{v})=\lim_{t\to \infty}f(\vec{v},t).
\]

We define $J_i\equiv d_i f-\frac{1}{2}\partial_i D_i f$ and
Eq.(\ref{eq:FP}) can be written as:
\[
\partial_t f(\vec{v},t)=-\sum_i \partial_i J_i.
\]
We obtain $f_{st}$ by solving  $J_i=0$ and we have:
\[
(d_i-\frac{1}{2}(\partial_iD_i))f_{st}=\frac{1}{2}D_i \partial_i f_{st}.
\]
The equation can be rewritten as
\[
Z_i\equiv \frac{(2d_i-\partial_i D_i)}{D_i}=\partial_i \ln f_{st}.
\]
We see  $Z_i$ should satisfy $\partial_j Z_i=\partial_i Z_j$.
A potential solution $f_{st}(\vec{v})=e^{-\phi(\vec{v})}$ of Eq.(\ref{eq:FP})
exists \cite{Gardiner:2009,Bakosi:2013} if $\phi$ satisfy
\[
-\partial_{i}\phi=\partial_{i}\ln f_{st}=\frac{2d_i-\partial_{i}D_i}{D_i}
\equiv Z_i
\]
From the constraint $\partial_j Z_i=\partial_i Z_j$,
we obtain 
\[
\frac{\theta_{2}}{v_i(1-v_i)}=\frac{\theta_{2}}{v_j(1-v_j)}.
\]
If we set $\theta_{2}=0$, the stationary solution
$f_{st}^{0}(\vec{v})$ becomes the direct product of the beta distribution
$f_{Beta}(v_i|a_i,b_i)$.
\[
f_{st}^{0}(\vec{v})=\prod_{i}f_{Beta}(v_i|a_i,b_i).
\]
The expectation value of $\vec{v}$ is $\vec{\mu}=(\mu_1,\mu_2,\cdots,\mu_I)$
and the covariance matrix $\Sigma^{0}$ of $\vec{v}$ is given by:
\[
\Sigma^{0}_{i,j}=\mbox{Cov}(\vec{v})_{i,j}
=\delta_{i,j}v_{i}(1-v_{j})\frac{1}{\theta_1+1}.
\]
When $\theta_1>>1$, the joint probability function $f_{st}^{0}(\vec{v})$
can be approximated by the multi-variate normal distribution as:
\[
\vec{v}\sim \mbox{N}_{I}(\vec{\mu},\Sigma^{0}).
\]

When $\theta_{2}\neq 0$, the potential solution does not exist.
$\vec{v}$ fluctuates around their equilibrium values
$\vec{v}^{*}$  and  $v_{i}^{*}$ is determined by the condition that $d_i=0$.
\[
v_{i}^{*}=\mu_i+(\theta_{2}|J(i)|/\theta_1)(v_{i,c}^{*}-v_{i}^{*}).
\]
Here, $v_{i,c}^{*}$ is the average value of $v_{j}^{*},j\in J(i)$.

When $\theta_2>>\theta_1$, $v_i^*\simeq v_{i,c}^{*}$ holds and 
$v_{i}^{*}$ is equal to the average value of $\mu_i$,
$\mu_{avg}\equiv\sum_i \mu_i/I$.
It is assumed that the fluctuation of $\vec{v}$ around $\mu_{avg}$
is small and approximates $D_{i}=2v_i(1-v_i)$ as
$D_i=D=2\mu_{avg}(1-\mu_{avg})$.
In this case,
the potential condition is satisfied.
We call the approximation the "Gaussian approximation", because
the Wright-Fisher diffusion eq. has a solution that can be
approximated as a Gaussian distribution.
We obtain $\ln f_{st}$ as:
\[
\ln f_{st} =\int^{\vec{v}}\sum_{i}Z_i dv_{i}
  =\frac{1}{\mu_{avg}(1-\mu_{avg})}
  \left(\theta_1\vec{\mu}\cdot \vec{v}-\mu_{avg}(1-\mu_{avg})
  \frac{1}{2}{}^{t}\vec{v}\Sigma^{-1}\vec{v}\right) 
  \]
The inverse of the covariance matrix $\Sigma^{-1}$ is
\[
\mu_{avg}(1-\mu_{avg})(\Sigma^{-1})_{i,j}=
\left\{
\begin{array}{cc}
  \theta_1+\theta_{2}|J(i)| & i=j \\
  -\theta_{2} & j \in J(i) \\
  0 & i\neq j,j \notin J(i)
\end{array}
\right.
\]
The multivariate normal approximation of $\vec{v}$ is
\begin{equation}
  \vec{v}\sim \mbox{N}_{I}(\vec{v}^{*},
  \Sigma) \label{eq:MN}.
\end{equation}

\section{\label{sec:simulation} Numerical study}
We then numerically verify the validity of the 
normal distribution approximation.
The conditional probability density function for $v_i$
with $v_{i,c}$ fixed is a beta distribution with the shape parameters
$a_{i}(v_{i,c})=\theta_1 \mu_i+\theta_{2}|J(i)|v_{i,c},
b_{i}(v_{i,c})=\theta_1 (1-\mu_i)+\theta_{2}|J(i)|(1-v_{i,c})$.
\[
v_{i}\sim \mbox{Beta}(a_{i}(v_{i,c}),b_{i}(v_{i,c})).
\]
We set the initial values for
$\vec{v}$ as $v_{i}\sim \mbox{Beta}(a_i,b_i)$
with $v_{i,c}=0.5$.
Thereafter, we choose a node $i$ at random and calculate the shape
parameters $a_{i}(v_{i}^{c}),b_{i}(v_{i}^{c})$ and generate new $v_{i}$
according to $v_{i}\sim \mbox{Beta}(a_{i}(v_{i}^{c}),b_{i}(v_{i}^{c}))$.
The process is repeated for $I$ times (1 MCS) and 
we obtain a sample $\vec{v}(1)$. The procedure is repeated
with the initial condition $\vec{v}(t),t=1,\cdots$ and 
we obtain a sample $\vec{v}(t+1)$.
The length (MCS) of the sample sequence $T$ is set as $10^{6}$.
In a 2D system, we set $T=2\times 10^5$.

\subsection{Two nodes ($I=2$) case}

\begin{figure}[htbp]
\begin{center}
\begin{tabular}{cc}
\includegraphics[width=7cm]{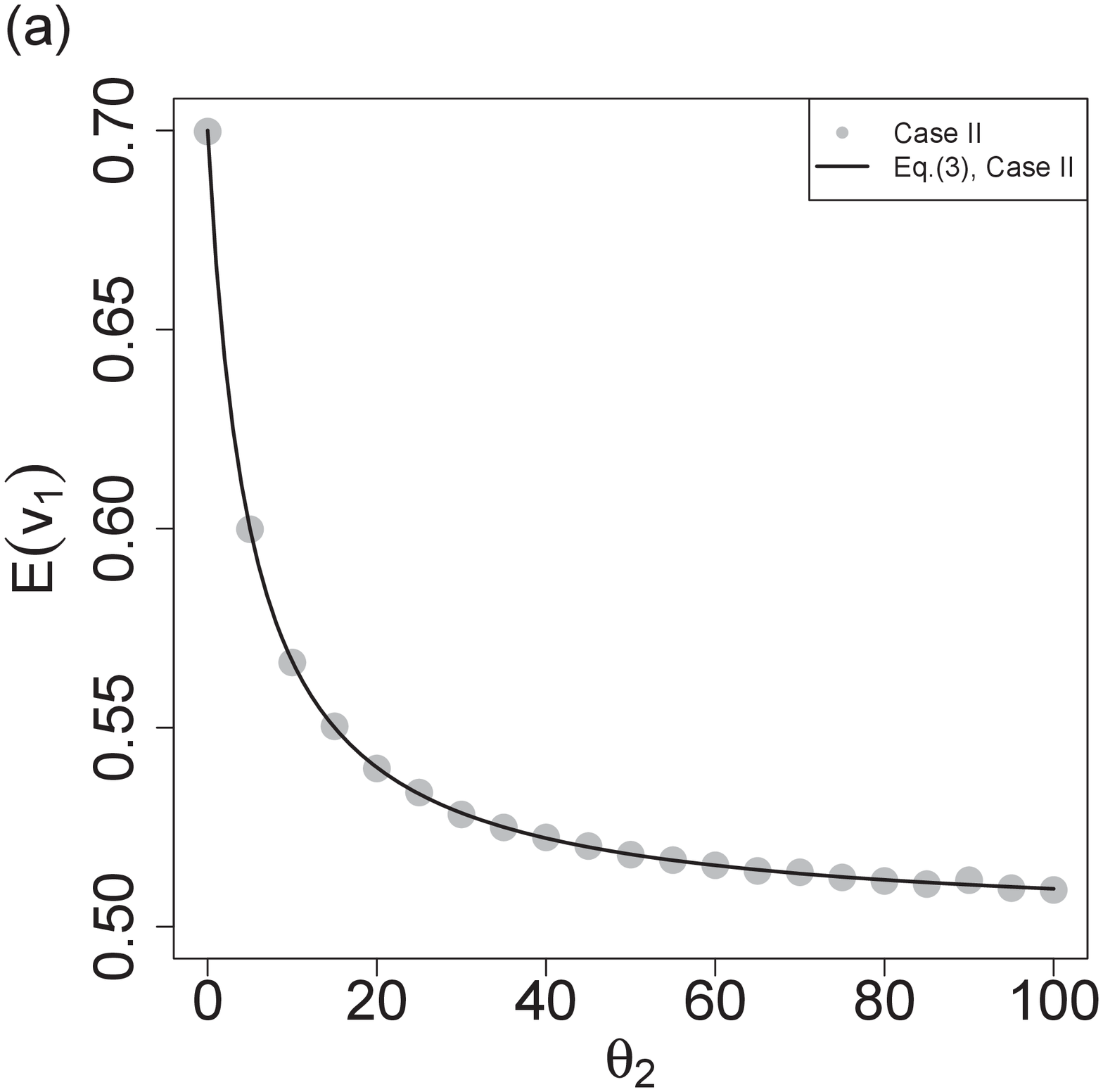}
&
\includegraphics[width=7cm]{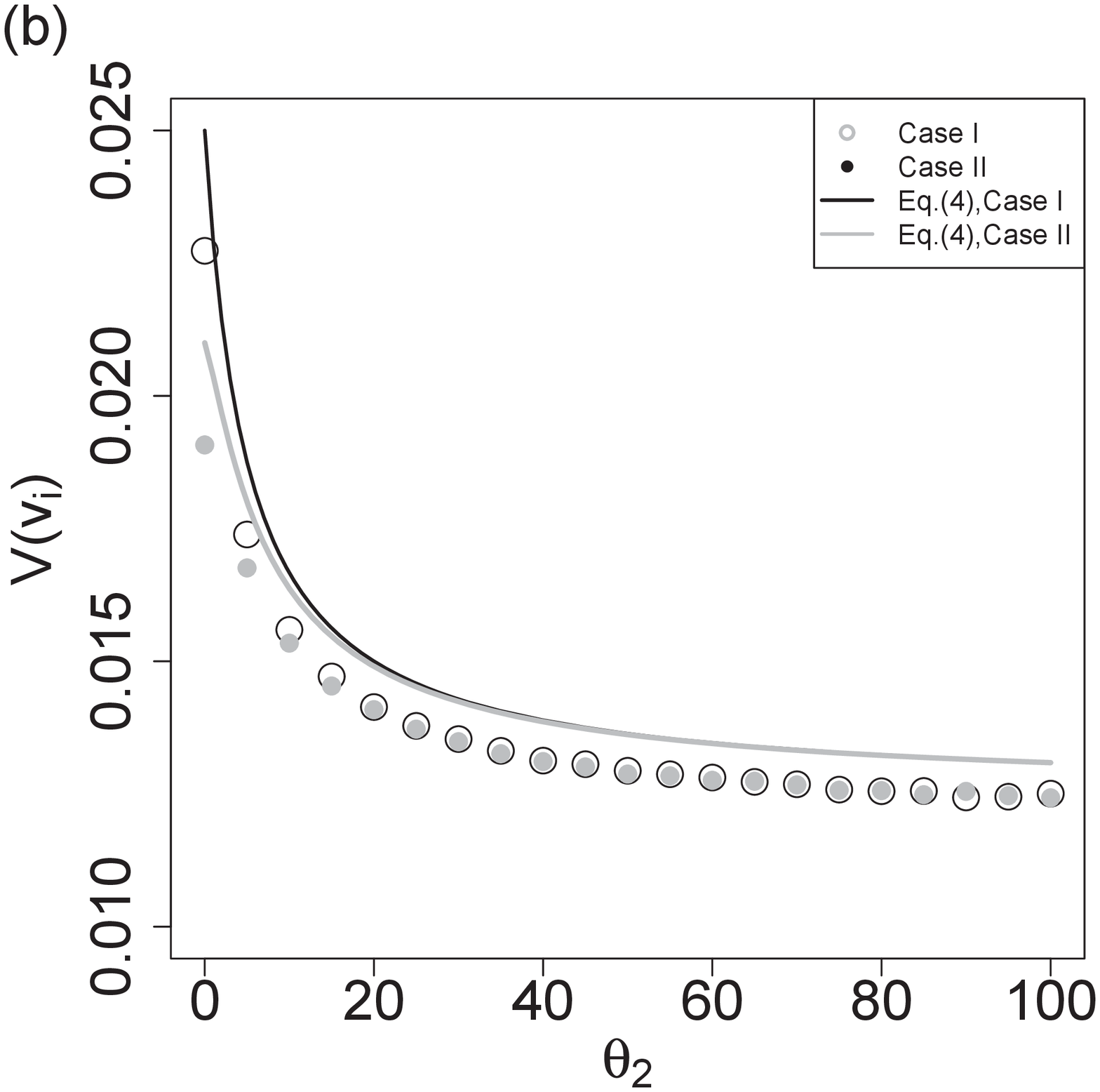}
\\
\includegraphics[width=7cm]{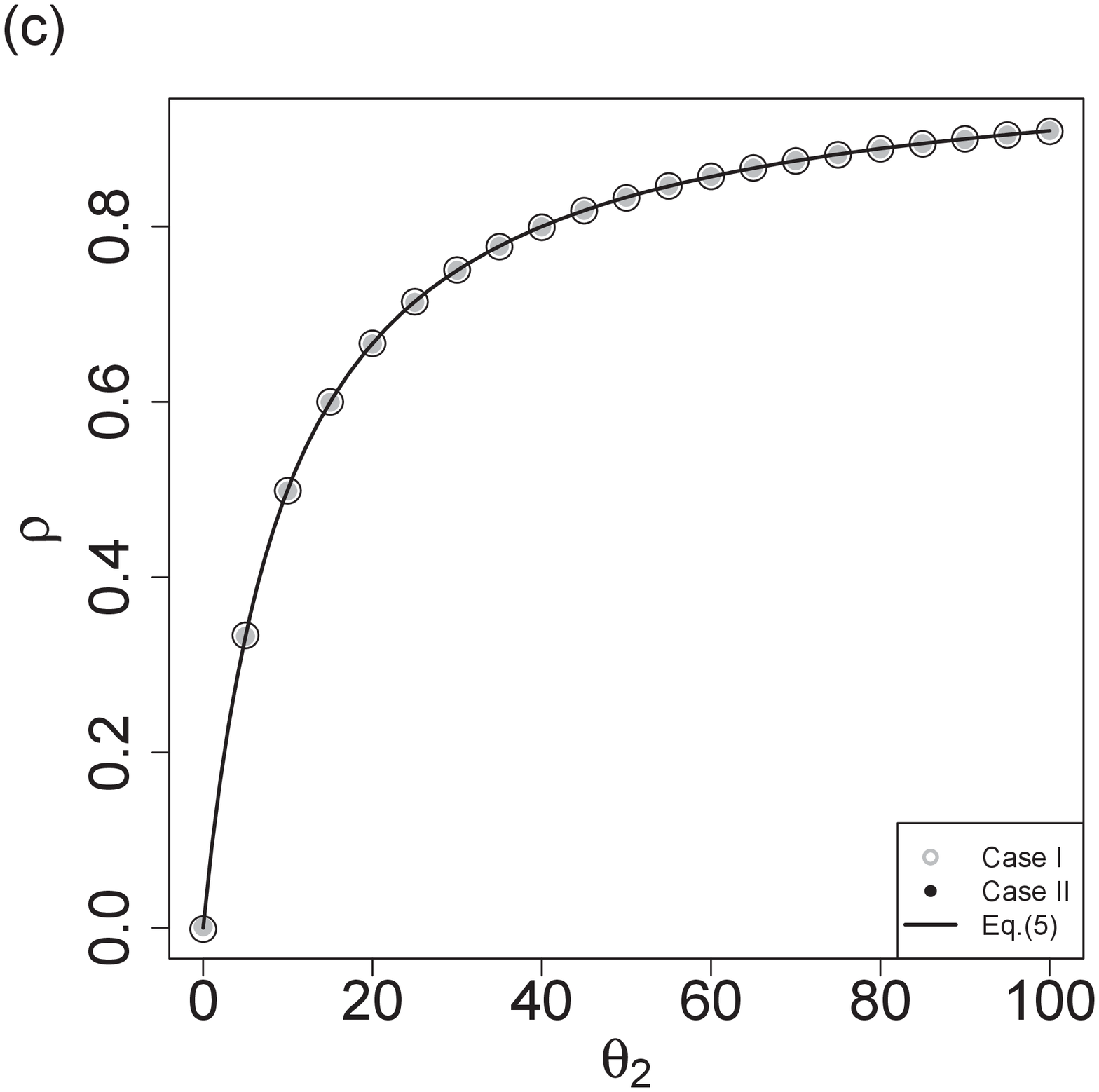}
\end{tabular}
\end{center}
\caption{\label{fig:4}
  (a) Plots of $\mbox{E}(v_1)$ vs. $\theta_2$ for case II. 
  Solid line plots of Eq.~(\ref{eq:Ev12}) vs. $\theta_2$.
  (b)  Plots of $\mbox{V}(v_1)$ vs. $\theta_2$ 
  for case I (gray circle), and case II (hollow black circle).
  The theoretical results for Eq.~(\ref{eq:Vvi})
  are plotted using solid curves in black and gray for case I and case II,
  respectively.
  (c)
 Plots of $\rho$ vs. $\theta_c$ 
 for case I (gray circle), case II (hollow black circle)
 and the theoretical results (solid curves)
    of Eq.~(\ref{eq:rho_12}).   
}
\end{figure}

At first, we consider the $I=2$ case.
We adopt $\mu_1,\mu_2$ so that $\mu_{avg}=0.5$. 
We then set $\theta_1=10$ and
$a_{i}=b_{i}=5,\mu_i=0.5$ in case I. In case II, we set $\theta_1=10$ and
$a_{1}=b_{2}=7,a_{2}=b_{1}=3,\mu_1=0.7,\mu_2=0.3$.
Based on the symmetry of the system,
$v_{i}^{*}$ is estimated as:
\begin{equation}
v_{1,2}^{*}=\frac{1}{2}\frac{a_1+a_2}{\theta_1}\pm
\frac{a_1-a_2}{\theta_1+2\theta_2}  \label{eq:Ev12}
\end{equation}
The variance of $v_i$ is:
\begin{equation}
  \mbox{V}(v_i)=\mu_{avg}(1-\mu_{avg})
  \frac{\theta_1+\theta_2}{\theta_1^2+2\theta_1\theta_2}
  \label{eq:Vvi}
\end{equation}
The correlation coefficient $\rho$ of $v_1$ and $v_2$ is:
\begin{equation}
\rho \equiv \frac{\mbox{Cov}(v_1,v_2)}{\sqrt{\mbox{V}(v_1)\mbox{V}(v_2)}}
=\frac{\theta_2}{\theta_1+\theta_2} \label{eq:rho_12}.
\end{equation}

Figure \ref{fig:4} shows the results of the MC studies.
The numerical data are plotted with symbols and the Gaussian approximation
results are presented with lines. Figure \ref{fig:4}(a) shows
E$(v_i)$ and $v_{1}^{*}$in Eq.(\ref{eq:Ev12})  vs. $\theta_2$ for case II.
Figure \ref{fig:4}(b) shows V$(v_i)$ and Eq.(\ref{eq:Vvi}) vs. $\theta_2$.
Figure \ref{fig:4}(c) shows the correlation coefficient $\rho$ 
and Eq.(\ref{eq:rho_12}) vs. $\theta_2$. There is some discrepancy in the
estimation of the variance, which originates from the diffusion approximation.
We see that the Gaussian approximation works well.

\subsection{Lattice case}

Next, we investigate the 1D lattice and 2D square lattice cases.
We are interested in the $r$ dependence of the correlation.
We consider $L$ sites for a 1D lattice and $L\times L$
sites for a 2D lattice.
The periodic boundary condition is imposed in both cases.
The nodes are indexed by $i\in \{1,\cdots,L\}$ for the 1D lattice and
$(i,j), i,j \in \{1,\cdots,L\}$ for the 2D lattice, respectively.
Nodes are linked with their nearest neighbors and $|J(i)|=2(4)$
for a 1D (2D) lattice.
We set $\mu_i=\mu_{(i,j)}=\mu_{avg}=1/2$ and $\theta_1=10$.

\begin{figure}[htbp]
\begin{center}
\begin{tabular}{cc}
\includegraphics[width=7cm]{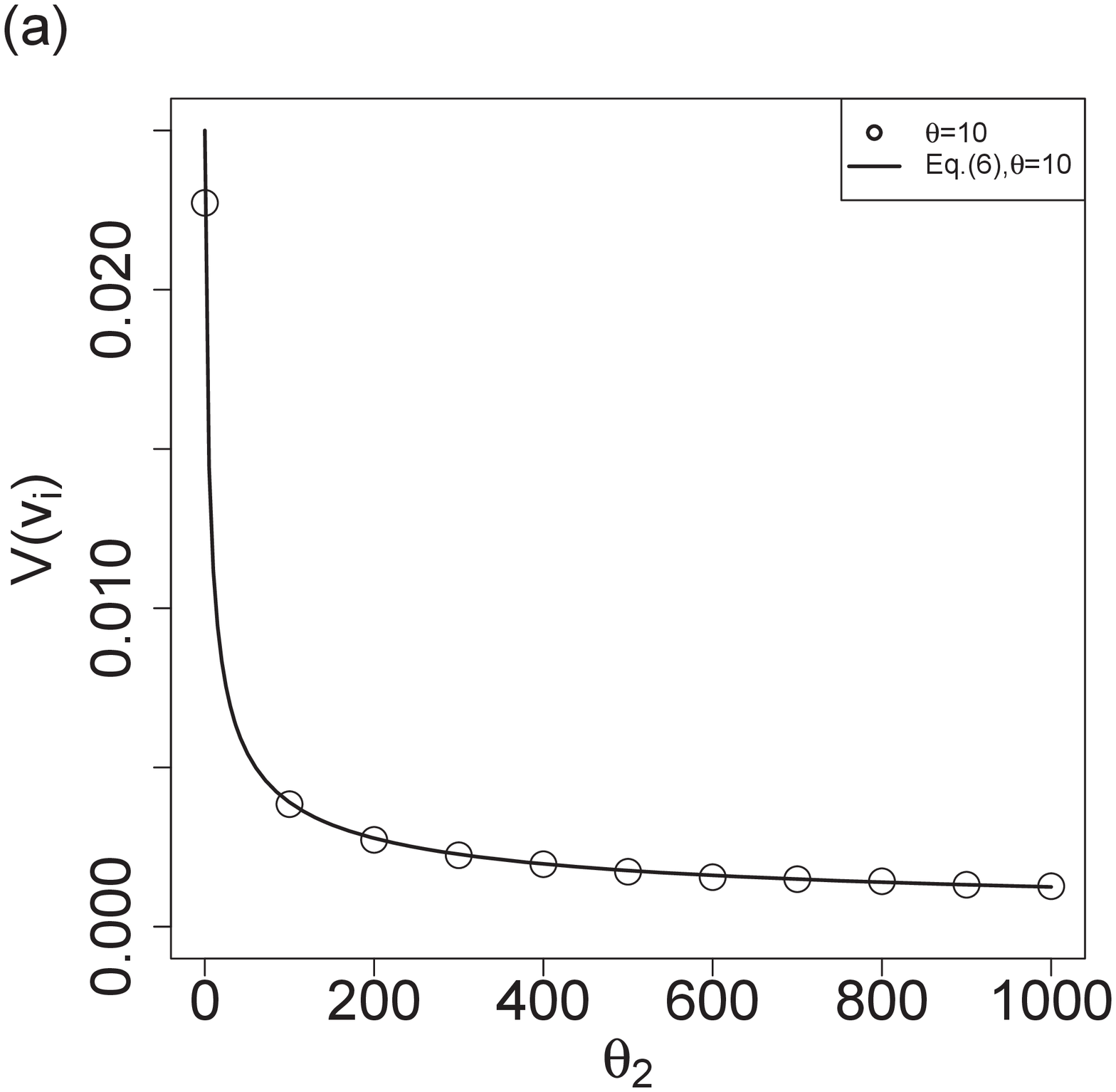}
&  
\includegraphics[width=7cm]{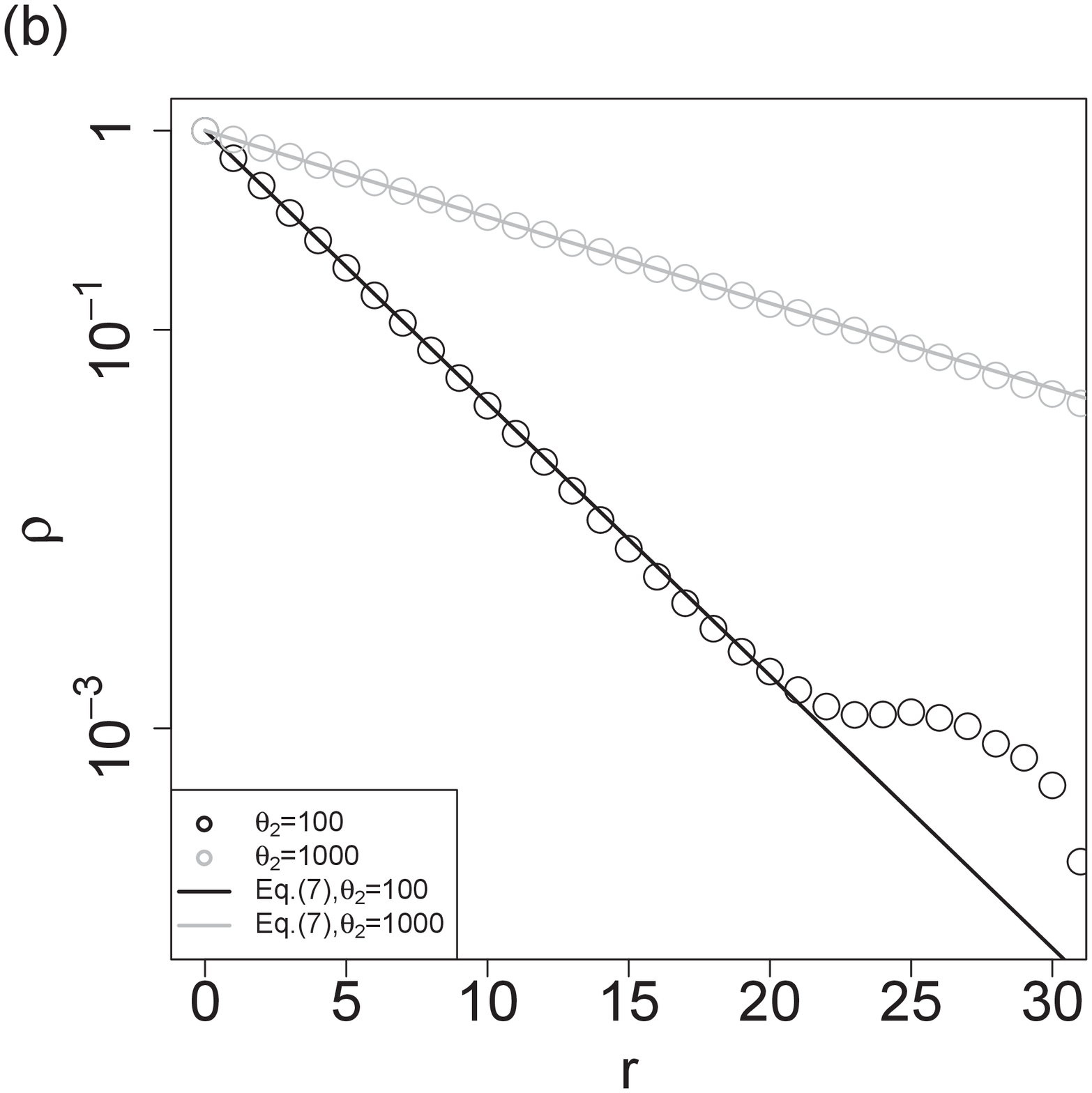}
\\  
\includegraphics[width=7cm]{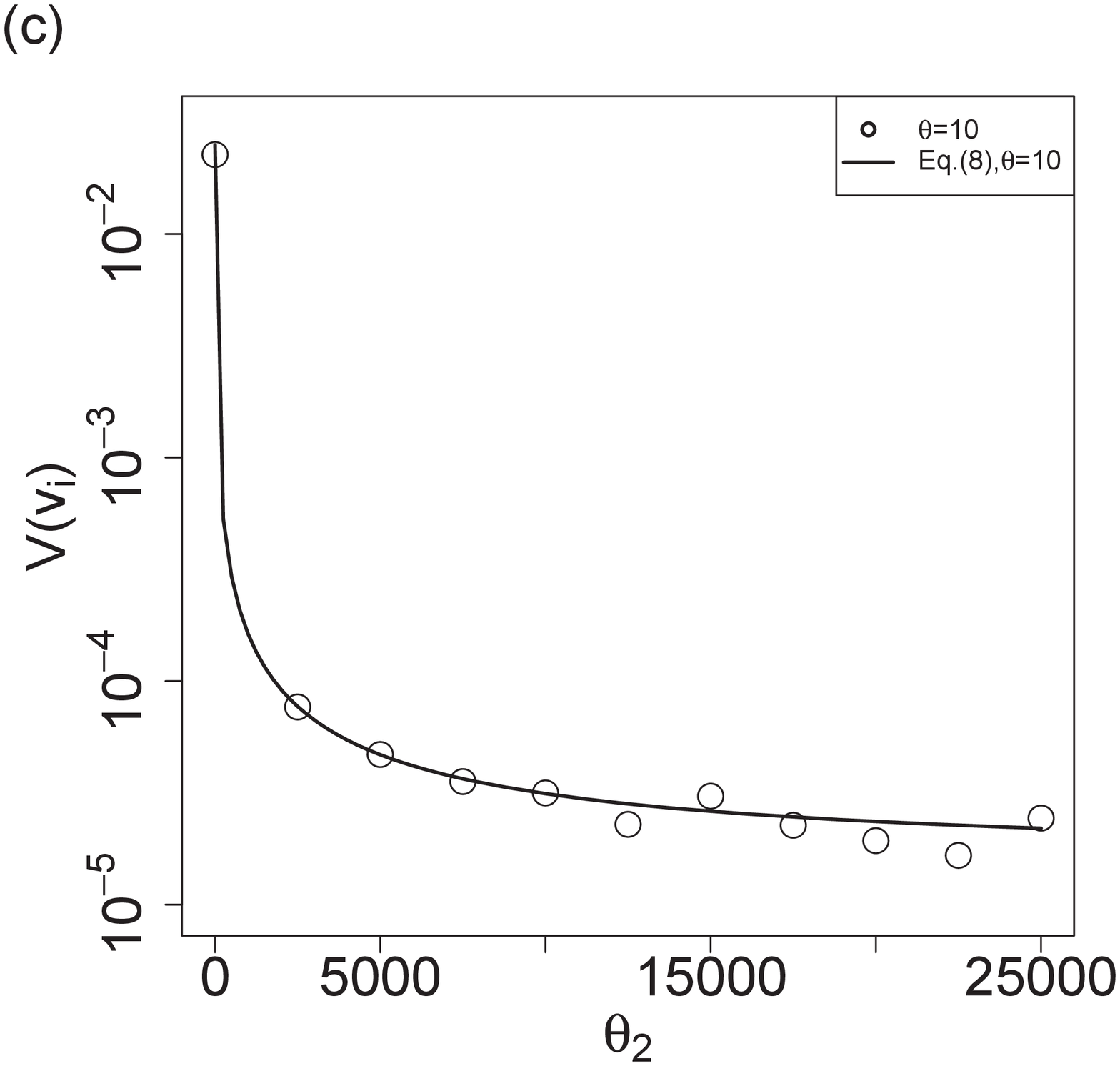} 
&
\includegraphics[width=7cm]{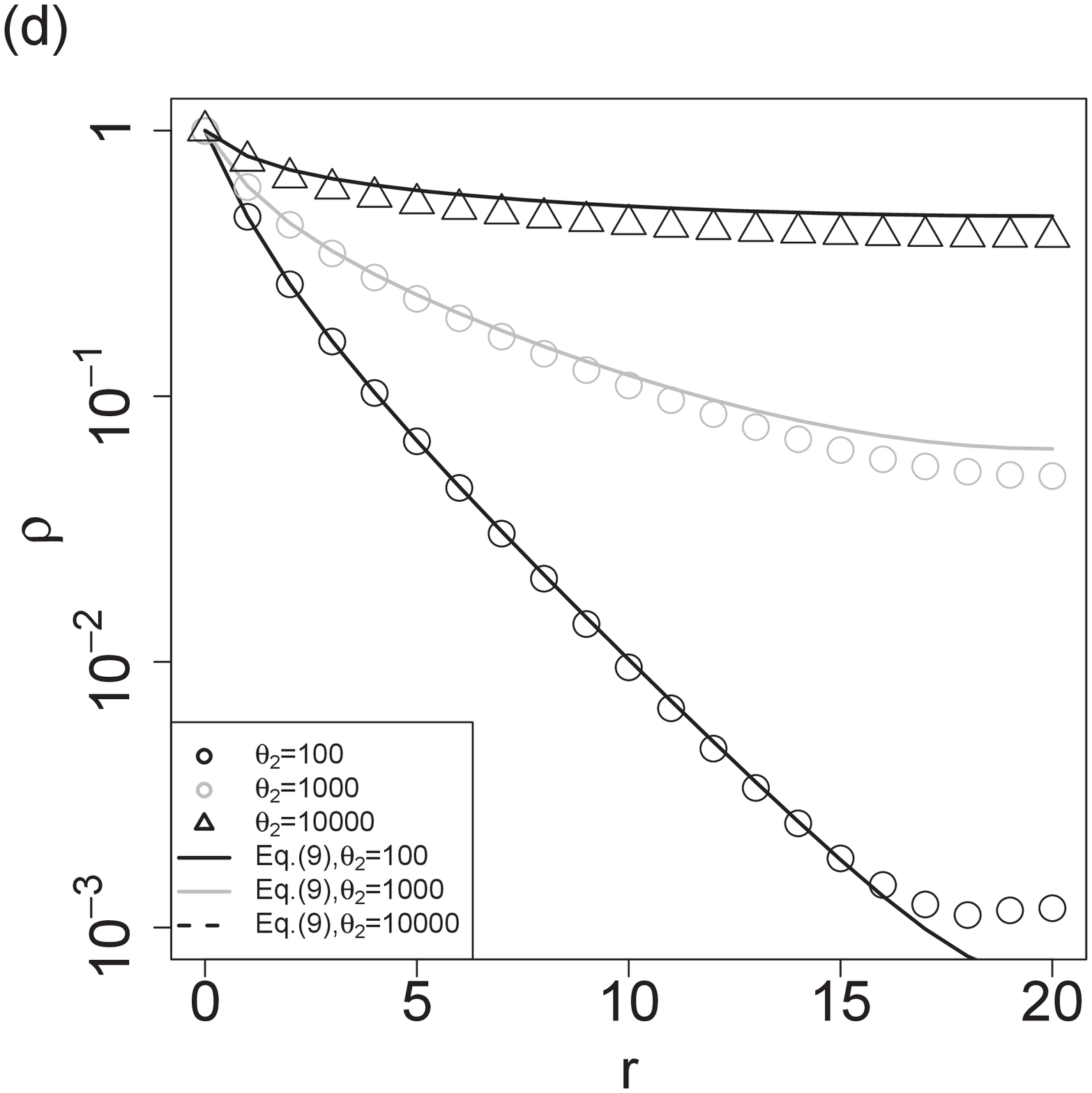}

\end{tabular}
\end{center}
\caption{\label{fig:Lattice}
  (a) Plots of $\mbox{V}(v_i)$ vs. $\theta_2$ for a 1D lattice with $L=10^2$.
  (b) Plots of $\rho(r)$ vs. $r$ for a 1D lattice with $L=10^2$.
  (c) Plots of $\mbox{V}(v_{(i,j)})$ vs. $\theta_2$ for a 2D lattice with $L=40$.
  (d) Semi-logarithmic plot of $\rho(r)$ vs. $r$ for a 2D with $L=40$.
  We adopt $\theta=10$ and $\mu_i=\mu_{(i,j)}=\mu_{avg}=1/2$. In (a) and (c), 
  solid line plots Eq.~(\ref{eq:Vvi_1D}) vs. $\theta_2$ for a 1D
  and Eq.~(\ref{eq:Vvi_2D}) vs. $\theta_2$ for a 2D.
  We adopt $\theta_{2}=10^2$ and $10^{3}$ for a 1D lattice in (b) 
  and $\theta_{2}=10^2,10^3$ and $\theta_{2}=10^4$
  for a 2D lattice in (d). 
  In (b) and (d), we plot the theoretical results from Eq.~(\ref{eq:rho_1D}) for 1D
  and those from Eq.~(\ref{eq:rho_2D}) for 2D by solid and broken curves.
}
\end{figure}	

$\vec{v}$ obeys a multi-variate normal
distribution for 1D lattice case.
The inverse of the covariance matrix $\Sigma^{-1}$ for the 1D lattice
is: 
\[
(\Sigma^{-1})_{i,j}=(\theta_1+2\theta_{2})\delta_{i,j}+
\theta_{2}(\delta_{i+1,j}+\delta_{i-1,j}).
\]
For a 2D lattice, the inverse of the covariance matrix $\Sigma^{-1}$
is: 
\[
(\Sigma^{-1})_{(i,j),{k,l}}=(\theta+4\theta_{2})\delta_{i,k}\delta_{j,l}
+\theta_{2}(\delta_{i+1,k}\delta_{j,l}
+\delta_{i-1,k}\delta_{j,l}+\delta_{i,k}\delta_{j+1,l}+\delta_{i,k}
\delta_{j-1,l}).
\]

The variance of $v_{i}$ is given by:
\begin{equation}
\mbox{V}(v_i)=\mu_{avg}(1-\mu_{avg})\Sigma_{i,i} \label{eq:Vvi_1D}.  
\end{equation}  
The correlation between $v_{i}$ and $v_{1+r}$ is:
\begin{equation}
\rho(r) =\Sigma_{1,1+r}/\Sigma_{1,1}.  
\label{eq:rho_1D}  
\end{equation} 
For a 2D lattice case, we obtain similar equations by replacing
$\Sigma_{i,j}$ with $\Sigma_{(i,j),(k,l)}$.
\begin{equation}
\mbox{V}(v_{(i,j)})=\mu_{avg}(1-\mu_{avg})\Sigma_{(i,j),(k,l)} \label{eq:Vvi_2D}.  
\end{equation}  
The correlation between $v_{(i,j)}$ and $v_{(i+r,j)}$ is
\begin{equation}
\rho(r) =\Sigma_{(1,1),(1+r,1)}/\Sigma_{(1,1),(1,1)}.  
\label{eq:rho_2D}  
\end{equation}

Figure \ref{fig:Lattice} shows a comparison of the MC data with
the results of the
Gaussian approximation. It is evident that the multivariate normal
distribution describes the joint probability function of $\vec{v}$
quite well.
Furthermore, we can confirm that the $r$ dependence of the
correlation decays exponentially with $r$ for the 1D lattice case.
For the 2D case, 
we also observe the exponential decay for
$\theta_2=10^2$.  For large $\theta_2=10^3,10^4$,
the $r$ dependence does not obey an exponential decay.
The correlation length
becomes comparable with the system size $L=40$ and the
exponential decay is not observed for the limited system size.

\subsection{U.S. county network case}

Here, we calibrate the model parameters $\theta_1,\theta_2$ using the U.S.
presidential election data in Section \ref{sec:data}
and the Gaussian approximation of the model.
We construct an artificial county network where 3105 counties constitute
nodes of the network and the counties with
their nearest $z$ neighbors are connected as links.
Here, $z$ neighbors are determined based 
on the geodesic distance of the separation of the centroids.
If $j\in J(i)$  and $i \notin J(j)$, we add $i$ in $J(j)$.
The number of neighbors depend on $i$.
We adopt $z\in \{3,4,5\}$ and all the nodes are included in the largest
components.
We then set $\theta_1,\theta_2$
so that SD of $v_i$ is approximately 8\% and
the correlation $\rho$ between the nearest
neighbor counties becomes approximately 83\%, which are the empirical values
in Section \ref{sec:data}. We adopt $\theta_1=\{0.044,0.034,0.03\}$
and $\theta_2=\{73,50,40\}$ for $z=\{3,4,5\}$, respectively.
Given that $\theta_2>>\theta_1$, the equilibrium values $v_{i}^{*}$ are almost equal.
This suggests that the cultural field
cannot be encoded in the model parameters $\mu_i$.
This point is discussed in the last section. Here, we adopt $\mu_i=1/2$.

\begin{figure}[htbp]
\begin{center}
\includegraphics[width=14cm]{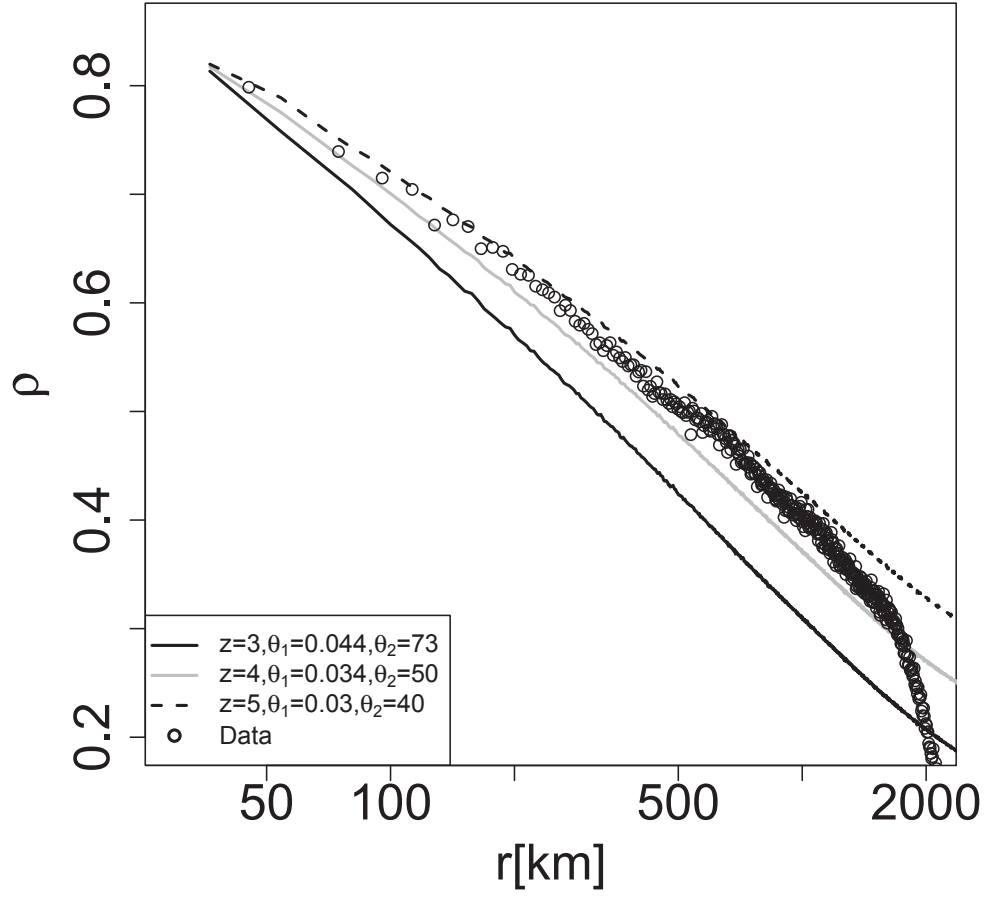}
\end{center}
\caption{\label{fig:6}  
  Semi-logarithmic plots of the correlation of $v_i$ (model) vs. $r$.
  We adopt $\theta_1=\{0.044,0.034,0.03\}$ and $\theta_2=\{73,50,40\}$ for $z=\{3,4,5\}$, respectively.
  The Gaussian approximation is used to estimate the correlation of the models.
  The symbols ($\circ$) is used to plot the correlation of $\Delta v_{d}$ vs. $r$.
}
\end{figure}

Figure \ref{fig:6} shows the results.
The correlation $\rho$ of $\vec{v}(t)$
is plotted as a function of $r$ for the three cases $z=3$
(solid black),$z=4$(solid, gray)
and $z=5$(broken black). 
When we set $\mu_{avg}=0.5$, $v_{i}^{*}=0.5$ and 
the correlation of
$\vec{v}(t)$ are the same as that of $\Delta \vec{v}(t)$. 
They start from the same value of 83\%
of the nearest neighbor correlation and decay monotonically with $r$.
As $z$ increases, the decay rate becomes small and the model shows a longer
spatial correlation. We also plot the empirical results
of the correlation of $\Delta \vec{v}_{d}(t)$
as a function of $r$ using
the symbols $\circ$. The $z=5$ case best fits the empirical
behavior of the correlation of $\Delta \vec{v}_{d}(t)$
among the three cases.

\section{\label{sec:Con} Conclusions}
In this report, we study the fluctuation of vote share in US
presidential election data. Compared with
the temporal average of
the vote shares 
in each county, the fluctuation shows a stronger
and long-range correlation.
In order to describe the behavior, we propose
a voter model on networks.
There are many voters in each node
and they choose another
voter at random and
copy the another voter's  choice as in the case of the voter model.
Another voter is selected from
the same node where the voter lives or
from a neighboring node.
Each node has an intrinsic parameter $\vec{\mu}$ that determines
the preference for an option. In addition, $\theta_1$ and $\theta_2$
incorporate the influence from the voters in the nodes where
the voters live and
from the voters
in the linked nodes, respectively. 
We derive the multivariate Wright-Fisher diffusion equation for the
joint probability density function (pdf) of the vote shares.
The pdf is a multivariate generalization of the beta distribution.
We approximate the pdf using the multivariate normal distribution
and estimate the variance and the correlation coefficient of the vote
shares. The results were then checked numerically.

There are a few unresolved problems for future study.
For example, the statistical modeling of elections and 
estimation of the model parameters,
$\vec{\mu},\theta_1$ and $\theta_2$ that can fit the
empirical nature of the election data need to be investigated further.
The estimation should be compatible with the 
long-range nature
of the correlation with distance $r$.
We think it is necessary to generalize 
the model by incorporating several types of voters.
As the equilibrium values $v_{i}^{*}$ becomes approximately equal
 to $\mu_{avg}$ when $\theta_2>>\theta_1$,  $\{v_{d}(i)\}$ cannot
 be encoded in $\{v_{i}^{*}\}$. 
The SIRM model avoids this
 problem by introducing a noise term\cite{Gracia:2014}.
 
 In order to realize $\vec{v}_{d}$ in our model, which is a
proxy of the cultural field, the assumption that all voters are
model voters is too simple.
Some voters do not change their choices even if they interact
with many other voters
of different choices.
This possibility was previously considered in the modeling
of Japan's parliament election with three political
parties\cite{Sano:2017}.
We assumed that there are two types of voters,
the fixed supporter of each political party and the floating voter.
The probability function then becomes the combination of the
multinomial distribution of the fixed supporter
and the Dirichlet distribution
of the floating voters. If we take into account the network structure
of the social influence, we have a combination of the multinomial
distribution and
a multivariate generalization of the Dirichlet distribution.
Using this idea, it is possible to incorporate $\vec{v}_{d}$
in the model.
It is also worthwhile to solve the Wright-Fisher diffusion equation for
the multi-variate beta distribution (Eq.\ref{eq:FP}).
Another type of multivariate beta distribution has been derived
for the inference in a statistical control process\cite{Adamski:2010}. 
The multivariate beta distribution for the voter model
on networks should be derived since it is the natural
multivariate extension
of a beta distribution based on the similarity with the multivariate normal
distribution.

\begin{acknowledgments}
This work is supported by JPSJ KAKENHI[Grant No. 17K00347].  
\end{acknowledgments}

\bibliography{myref201902}
\appendix

\section{Wright-Fisher diffusion for SIRM model}
In the SIRM model, voters who live in node $i$ move to
node $j$ for work\cite{Gracia:2014}.
They interact with other agents of node $i$ and those of
node $j$ in addition to the agents on the link $(i,j)$\cite{Michaud:2018}. 
We denote the number of voters on link $(i,j)$ as $N_{i,j}$ and
the number of votes for an option as $n_{i,j}$.
As in the social influence model on a network in the main text, we
introduce the parameters $\vec{\mu}$ that determine the
intrinsic preference for an option 
of node $i$. We also introduce $\theta_1$ and
$\theta_2$ which control
the variance of the vote share and 
correlation of the votes shares between nodes, respectively.
In the SIRM model, there is a parameter
$\alpha$ which controls the strength of social
influence from the node where a voter lives and works.
$N_{i\cdot}\equiv \sum_{j}N_{i,j}$ and
$n_{i\cdot}\equiv\sum_{j}n_{i,j}$ indicate the number of voters who
live in node $i$ and the number of votes for an option attributed to them.
$N_{\cdot j}\equiv \sum_{i}N_{i,j}$ and
$n_{\cdot j}\equiv\sum_{i}n_{i,j}$ indicate the number of voters
who work in node $j$ and the number of votes for an option due to them.
Likewise, we also write
$v_{i\cdot},v_{\cdot j}$ which represents the vote shares of the voters
who live in node $i$ and those who work in node $j$.
We then write $N_{T}=\sum_{i,j}N_{i,j}$ for the total number of voters.

We write the probabilities for $n_{i,j}\to n_{i,j}+1$ and $n_{i,j}\to n_{i,j}-1$ as:
\begin{eqnarray}
P(n_{i,j}\to n_{i,j}+1)&=&\frac{N_{i,j}-n_{i,j}}{N_{T}}
\cdot \frac{n_{i,j}+\alpha(a_i+\theta_2 v_{i\cdot})
  +(1-\alpha)(a_j+\theta_2 v_{\cdot j})}{N_{i,j}-1+\theta_1+\theta_2} \nonumber \\
P(n_{i,j}\to n_{i,j}-1)&=&\frac{n_{i,j}}{N_{T}}
\cdot \frac{N_{i,j}-n_{i,j}+\alpha(b_i+\theta_2(1-v_{i\cdot}))
  +(1-\alpha)(b_j+\theta_2(1-v_{\cdot j}))}{N_{i,j}-1+\theta_1+\theta_2} \nonumber
\end{eqnarray}
Here, $a_i\equiv \theta_1 \mu_i,b_i=\theta_1(1-\mu_i)$ are defined as before.
The raising operator $R_{i,j}(\vec{v})$ and the lowering operator
$L_{i,j}(\vec{v})$ are then defined as:
\begin{eqnarray}
  R_{i,j}(\vec{v})&=&\frac{N_{i,j}}{N_{T}}(1-v_{i,j})
  \cdot \frac{v_{i,j}+\alpha(a_i+\theta_2 v_{i\cdot})/N_{i,j}+
    (1-\alpha)(a_j+\theta_2v_{\cdot j})/N_{i,j}}
    {1-1/N_{i,j}+\theta_1/N_{i,j}+\theta_2/N_{i,j}}
  \nonumber \\
  L_{i,j}(\vec{v})&=&\frac{N_{i,j}}{N_{T}}v_{i,j}
  \cdot
  \frac{v_{i,j}+\alpha(a_i+\theta_2 v_{i\cdot})/N_{i,j}+
    (1-\alpha)(a_j+\theta_2v _{\cdot j})/N_{i,j}}
        {1-1/N_{i,j}+\theta_1/N_{i,j}+\theta_2/N_{i,j}}  \nonumber
\end{eqnarray}  
The stochastic differential equation for $v_{i,j}$ is written as:
\[
d v_{i,j}=d_{i,j}dt+\sqrt{D_{i,j}}W_{i,j}(t)
\]
Here $W_{i,j}(t)$ is an iid Wiener process. The drift term $d_{i,j}$
and diffusion terms $D_{i,j}$ are
written as
\begin{eqnarray}
  d_{i,j}&=&\alpha a_i+(1-\alpha)a_j-\theta v_{i,j}+\theta_2(\alpha v_{i\cdot}
  +(1-\alpha)v_{\cdot j}),\\
  D_{i,j}&=& 2v_{i,j}(1-v_{i,j}).
\end{eqnarray}

\section{Voter model on complete graph and beta binomial distribution}
There are $N$ voters and the number of voters who vote
for an option is denoted as $n$. The probability for
$n\to n+1$ is written as,
\[
P(n\to n+1)=\frac{N-n}{N}\cdot \frac{n+a}{N-1+\theta}.
\]
The probability for $n\to n-1$ is:
\[
P(n\to n-1)=\frac{n}{N}\cdot \frac{(N-n)+b}{N-1+\theta}.
\]
Here, $\theta=a+b$. In the stationary state, the condition of
the detailed balance between
$P(n)$ and $P(n+1)$ is,
\[
P(n)\dot P(n\to n+1)=P(n+1)\cdot P(n+1\to n).
\]
We obtain the next recursive relation for $P(n)$ and $P(n+1)$,
\[
P(n+1)=\frac{(N-n)(n+a)}{(n+1)(N-(n+1)+b)}P(n).
\]
The solution with the normalization $\sum_n P(n)=1$ is
the beta binomial distribution.
\[
P(n)={}_{N}C_{n}\frac{(a)^{n}(b)^{N-n}}{(\theta)^{N}}
={}_{N}C_{n}\frac{B(n+a,N-n+b)}{B(a,b)}.
\]
Here, $(x)^{n}\equiv \prod_{k=1}^{n}(x+k-1)$
is the rising factorial. Using the definition of the beta function,
we write $P(n)$ as:
\[
P(n)={}_{N}C_{n}\int_{0}^{1}p^{n}(1-p)^{N-n}\frac{p^{a-1}(1-p)^{b-1}}{B(a,b)}dp.
\]
In the continuum limit, the
pdf for $v\equiv \lim_{N\to \infty}n/N$ is the beta distribution with shape
parameters $(a,b)$.
\[
\lim_{N\to \infty}N\cdot P(n=Nv)=f_{Beta}(v|a,b)\equiv \frac{v^{a-1}(1-v)^{b-1}}{B(a,b)}.
\]

\end{document}